\documentstyle[12pt]{article}

\begin{document}

\title{TESTING THE VECTOR CONDENSATE MODEL OF ELECTROWEAK
INTERACTIONS AT HIGH ENERGY HADRON COLLIDERS}

\author{G.  Cynolter, E.  Lendvai and G.  P\'ocsik \\
 Institute for Theoretical Physics, \\ E\"otv\"os Lorand
University, Budapest }

\date{}
\maketitle


\begin{abstract}

In the vector condensate model a doublet of vector fields
plays the role of the Higgs doublet of standard model and
the gauge symmetry is broken dynamically.  This results in
a theory surviving the test of radiative corrections
provided the new charged and neutral vector particles B
have masses of at least several hundred GeV's.  In this
note we show that while at the Tevatron the heavy
B-particle production is too low, at LHC the yield is large
and, for instance, the inclusive cross section of $B^+ B^-$
pairs is 51.5 (15.3) fb at $\sqrt{s}=14$ TeV,
$m_B=400(500)$ GeV.

\end{abstract}

The vector condensate model employs the Lagrangian of the
standard model of electroweak interactions but the usual
scalar doublet is replaced by a doublet of vector fields
$B_{\mu}$ [1] whose neutral component forms a condensate
breaking the gauge symmetry dynamically and providing
nonvanishing $W,Z$ and vanishing photon masses.  A quartic
self-coupling of $B_{\mu}$ gives masses to the B-particles,
as well as the interaction of B-pairs with fermions gives
rise to fermion masses and embeds the Kobayashi-Maskawa
mechanism [1].

Looking at the S parameter it follows that the B-particles
are heavy, at $\Lambda=1$ TeV the threshold is about $m_0
\simeq 400-500 $ GeV for $B^0$ and $m_+ \simeq 200-350$ GeV
for $B^+$ [2].  Higher $\Lambda$ attracts higher minimum
masses.  The allowed regions by S are tightened by the T
parameter [3].  For example at $\Lambda=1 $ TeV, $m_0=400$
GeV the threshold for $m_+$ is increased to 630 GeV.  In
general, to each momentum scale there exists a range of B
masses where the radiative corrections are suitably small
by partial cancellations.  This is why $\Lambda$ cannot be
too large as compared to B masses.

Recently it has been shown that pairs of heavy B--particles
are copiously produced at high energy linear $e^+ e^-$
colliders [4].  In the present note we study the production
of B--particles at the Tevatron and LHC.  We show that
producing heavy B-particles at LHC is very favourable
having a large cross section while at Tevatron energy the
production cross section cannot exceed (0.01-0.02) fb which
is far below the discovery limit.

Since fermions are coupled very weakly to B-pairs [1] in
the vector condensate model, producing B-pairs is expected
to be more considerable from virtual $ \gamma $ and Z
exchanges, that is we consider the Drell-Yan mechanism [5],
$ p ^(\overline{p}^) \rightarrow B \overline{B} +X $ via
quark-antiquark annihilation.

The Drell-Yan cross section for the above hadronic
collisions can be written as [5,6] 
\begin{eqnarray} \sigma(
 p ^(\overline{p}^) \rightarrow B \overline{B} +X)=
\int_{\tau_0}^1 \, d\tau \int_{\tau}^1 \, \frac{dx}{2x}
\sum_i \sigma(q_i \overline{q}_i \rightarrow B
\overline{B}) \cdot \nonumber \\ \left ( f_i^1(x,\hat{s})
f_{\bar i}^2(\tau/x, \hat{s})+ f_{\bar i}^1(x,\hat{s})
f_{i}^2 (\tau/x, \hat{s}) \right )
\end{eqnarray}
where $x$ and $\tau/x$ are the parton momentum fractions,
$\hat{s}=\tau s$ is the square of the centre of mass energy
of $q_i \bar q_i$, s is the same for the hadronic initial
state, $f^1_i (x, \hat{s} )$ means the number distribution
of $i$ quarks in hadron 1 at the scale $\hat{s}$ and the
sum runs over the quark flavours u,d,s,c.  In the
computation the MRS (G) fit program [7] was used for the
parton distributions.

The angle integrated, colour averaged annihilation cross section
$ \sigma(q_i \overline{q}_i \rightarrow B \overline{B} ) $
is calculated to lowest order in the gauge couplings, and
QCD corrections are neglected.  We hope this approximation
shows the order of magnitude of the cross section.
 For $B^0 \overline{B}^0$
final state the Z exchange is working, and $B^+ B^- $
 pairs appear via $\gamma+Z $ exchange.
This is because in the model the following Lagrangians
relevant to the annihilation process emerge [1]:

\begin{eqnarray} 
L \left( B^{0} \right)&=&{ig \over 2cos
 \theta_W } \partial^{\mu} B^{(0)\nu +} \left( Z_{ \mu}
 B_{\nu}^{(0)} - Z_{\nu} B^{(0)}_{\mu} \right) + h.c.,
 \nonumber \\ L \left( B^+ {B^-} Z
 \right)&=&-cos2\theta_W \cdot L \left(B^{(0)}
 \rightarrow B^{(+)} \right),
\end{eqnarray} 
where $B^{(0)}_{\mu} \left( B^{(+)}_{\mu} \right ) $ denotes the
field of neutral (charged) B--particles. 
At $q_j \overline{q}_jZ$-vertex the usual coupling 
$ig \gamma_{\mu} (g_{Vj}+g_{Aj} \gamma_5 )$ acts, here

\begin{eqnarray} \left.  \matrix{
g_{Vj}& = & \frac{1}{2}&-\frac{4}{3} sin^2 \theta_W, \cr & & & \cr
g_{Aj}& = &{ \frac{1}{2} }& } \right \} j=u,c \nonumber \\ \\ \left.
\matrix{ g_{Vj}& = &-\frac{1}{2}&+\frac{2}{3} sin^2 \theta_W,
\cr & & & \cr g_{Aj}&=&-\frac{1}{2}& } \right \} j=d,s  \ .\nonumber
\end{eqnarray}

From (2) we get for $B^0 \overline{B}^0$ final states
\begin{eqnarray} 
\sigma(q_i \overline{q}_i \rightarrow B^0
\overline{B}^0)= \frac{1}{3} \frac{1}{256 \pi} \left (
\frac{g}{cos \theta_W} \right )^4 ( g_{Vi}^2+g_{Ai}^2)
\left( 1-\frac{4 m_0^2}{\hat{s}} \right)^{3/2} 
\frac{\hat{s} +8 m_0^2 }{4 m_0^4} .  
\end{eqnarray}

This is decreasing at high, increasing $m_0$ and for
$\hat{s} \gg 4 m_0^2$ it is proportional to
$\hat{s}/m_0^4$ reflecting that the Lagrangian (2) is
coming form a nonrenormalisable, effective model.  The
cross section of $B^+ B^-$ pairs can be expressed by (4) in
the following way 
\begin{eqnarray} 
\sigma(q_i
\overline{q}_i \rightarrow B^+ B^-)= \sigma(q_i
\overline{q}_i \rightarrow B^0 \overline{B}^0; m_0
\rightarrow m_+) \cdot \frac{1}{g_{Vi}^2+g_{Ai}^2} \cdot
\nonumber \\ \left [ (g_{Vi}^2+g_{Ai}^2 )cos^2 2 \theta_W+ 2
Q_{q_i} g_{Vi} sin^2 2 \theta_W cos 2 \theta_W + 4
Q_{q_i}^2 sin^2 2\theta_W \right].
\end{eqnarray}
The individual terms are due to $Z$ exchange, $\gamma-Z$
interference and $\gamma$ exchange. 

 We have calculated various distributions of
$B \overline{B}$ pairs for $p \overline{p}$
collisions at $\sqrt{s}=1.8 $ TeV and for $pp$ collisions
at $\sqrt{s}=14$ TeV, assuming $m_B=400,500,600$ GeV.
Typically, at the Tevatron no notable result can be
presented, however, increasing the energy to LHC, we get
sizable cross sections.  Also the yield of $B^+ B^-$ is
larger than that of $B^0 \overline{B}^0$.  As an example we
show in Fig.1 the $\tau$-distribution of $B^+ B^-$ pairs
at LHC, $m_+=400$ GeV.  $\frac{d \sigma}{d \tau}$ is
sharply peaked after threshold ($4m_+^2/s$) and
insignificant from about $\tau=0.1$, that is for higher
invariant masses of $B^+B^-.$ Fig.2 shows
$\frac{\partial^2 \sigma}{\partial p_T \partial y }
\vert_{y=0}$ as the function of the transverse momentum
$p_T$ of $B^+$ at vanishing rapidity y of $B^+ B^-$ for
LHC, $m_+=400$ GeV.

For the total cross section (1) we obtain 
\begin{eqnarray}
& \sigma_{\rm Tev}&=0.020 (0.0145) fb \, \hbox{ for } B^+B^-
(B^0 \overline{B}^0), m_B=400 \hbox{GeV}, \nonumber \\
&\sigma_{LHC}&=51.46; 15.30; 5.65 fb \; (44.35; 13.54; 4.90
fb) \\ & & \hbox{ for } B^+B^- (B^0 \overline{B}^0),
m_B=0.4; 0.5; 0.6 \hbox{TeV}.  \nonumber 
\end{eqnarray}

For instance, at an expected integrated luminosity of $10^5
{\rm pb}^{-1}$ one gets about 5100 $B^+ B^-$ pairs of
$m_+=0.4 $ TeV at LHC per annum.

In conclusion, we have shown that heavy B--particle pairs
have a large inclusive cross section due to $q
\overline{q}$ annihilation at LHC in a suitable mass range
making the detection of $B^{+,0}$ at LHC possible.
\vspace*{.5cm}

This work is supported in part by OTKA I/7, No.  16248.
\pagebreak

\pagebreak

{\Large \bf Figure Captions}
\vspace*{1cm}

Fig.  1:  The differential $\tau$-distribution of $B^+B^-$
pairs at LHC,$m_+=400$ GeV.

\vspace*{.5cm}

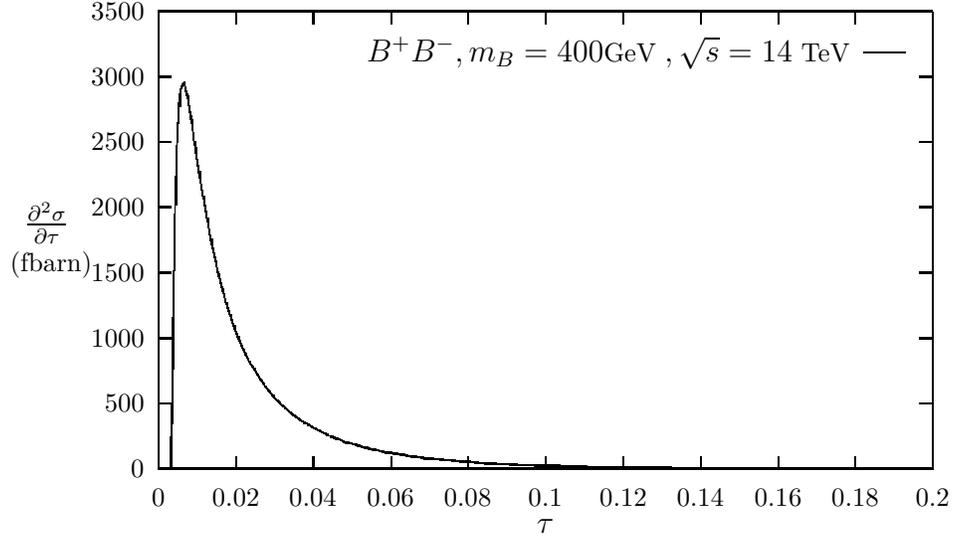
Fig.  2:  $\frac{\partial^2 \sigma}{\partial p_T \partial y
 } \vert_{y=0}$ at LHC, $m_+=400$ GeV.  $p_T$ denotes the
 transverse momentum of $B^+$ and $y$ is the rapidity of
 $B^+B^-$.

\pagebreak

\setlength{\unitlength}{0.240900pt}
\ifx\plotpoint\undefined\newsavebox{\plotpoint}\fi
\sbox{\plotpoint}{\rule[-0.200pt]{0.400pt}{0.400pt}}%
\begin{picture}(1500,900)(0,0)
\font\gnuplot=cmr10 at 10pt
\gnuplot
\sbox{\plotpoint}{\rule[-0.200pt]{0.400pt}{0.400pt}}%
\put(220.0,113.0){\rule[-0.200pt]{292.934pt}{0.400pt}}
\put(220.0,113.0){\rule[-0.200pt]{0.400pt}{173.207pt}}
\put(220.0,113.0){\rule[-0.200pt]{4.818pt}{0.400pt}}
\put(198,113){\makebox(0,0)[r]{0}}
\put(1416.0,113.0){\rule[-0.200pt]{4.818pt}{0.400pt}}
\put(220.0,216.0){\rule[-0.200pt]{4.818pt}{0.400pt}}
\put(198,216){\makebox(0,0)[r]{500}}
\put(1416.0,216.0){\rule[-0.200pt]{4.818pt}{0.400pt}}
\put(220.0,318.0){\rule[-0.200pt]{4.818pt}{0.400pt}}
\put(198,318){\makebox(0,0)[r]{1000}}
\put(1416.0,318.0){\rule[-0.200pt]{4.818pt}{0.400pt}}
\put(220.0,421.0){\rule[-0.200pt]{4.818pt}{0.400pt}}
\put(198,421){\makebox(0,0)[r]{1500}}
\put(1416.0,421.0){\rule[-0.200pt]{4.818pt}{0.400pt}}
\put(220.0,524.0){\rule[-0.200pt]{4.818pt}{0.400pt}}
\put(198,524){\makebox(0,0)[r]{2000}}
\put(1416.0,524.0){\rule[-0.200pt]{4.818pt}{0.400pt}}
\put(220.0,627.0){\rule[-0.200pt]{4.818pt}{0.400pt}}
\put(198,627){\makebox(0,0)[r]{2500}}
\put(1416.0,627.0){\rule[-0.200pt]{4.818pt}{0.400pt}}
\put(220.0,729.0){\rule[-0.200pt]{4.818pt}{0.400pt}}
\put(198,729){\makebox(0,0)[r]{3000}}
\put(1416.0,729.0){\rule[-0.200pt]{4.818pt}{0.400pt}}
\put(220.0,832.0){\rule[-0.200pt]{4.818pt}{0.400pt}}
\put(198,832){\makebox(0,0)[r]{3500}}
\put(1416.0,832.0){\rule[-0.200pt]{4.818pt}{0.400pt}}
\put(220.0,113.0){\rule[-0.200pt]{0.400pt}{4.818pt}}
\put(220,68){\makebox(0,0){0}}
\put(220.0,812.0){\rule[-0.200pt]{0.400pt}{4.818pt}}
\put(342.0,113.0){\rule[-0.200pt]{0.400pt}{4.818pt}}
\put(342,68){\makebox(0,0){0.02}}
\put(342.0,812.0){\rule[-0.200pt]{0.400pt}{4.818pt}}
\put(463.0,113.0){\rule[-0.200pt]{0.400pt}{4.818pt}}
\put(463,68){\makebox(0,0){0.04}}
\put(463.0,812.0){\rule[-0.200pt]{0.400pt}{4.818pt}}
\put(585.0,113.0){\rule[-0.200pt]{0.400pt}{4.818pt}}
\put(585,68){\makebox(0,0){0.06}}
\put(585.0,812.0){\rule[-0.200pt]{0.400pt}{4.818pt}}
\put(706.0,113.0){\rule[-0.200pt]{0.400pt}{4.818pt}}
\put(706,68){\makebox(0,0){0.08}}
\put(706.0,812.0){\rule[-0.200pt]{0.400pt}{4.818pt}}
\put(828.0,113.0){\rule[-0.200pt]{0.400pt}{4.818pt}}
\put(828,68){\makebox(0,0){0.1}}
\put(828.0,812.0){\rule[-0.200pt]{0.400pt}{4.818pt}}
\put(950.0,113.0){\rule[-0.200pt]{0.400pt}{4.818pt}}
\put(950,68){\makebox(0,0){0.12}}
\put(950.0,812.0){\rule[-0.200pt]{0.400pt}{4.818pt}}
\put(1071.0,113.0){\rule[-0.200pt]{0.400pt}{4.818pt}}
\put(1071,68){\makebox(0,0){0.14}}
\put(1071.0,812.0){\rule[-0.200pt]{0.400pt}{4.818pt}}
\put(1193.0,113.0){\rule[-0.200pt]{0.400pt}{4.818pt}}
\put(1193,68){\makebox(0,0){0.16}}
\put(1193.0,812.0){\rule[-0.200pt]{0.400pt}{4.818pt}}
\put(1314.0,113.0){\rule[-0.200pt]{0.400pt}{4.818pt}}
\put(1314,68){\makebox(0,0){0.18}}
\put(1314.0,812.0){\rule[-0.200pt]{0.400pt}{4.818pt}}
\put(1436.0,113.0){\rule[-0.200pt]{0.400pt}{4.818pt}}
\put(1436,68){\makebox(0,0){0.2}}
\put(1436.0,812.0){\rule[-0.200pt]{0.400pt}{4.818pt}}
\put(220.0,113.0){\rule[-0.200pt]{292.934pt}{0.400pt}}
\put(1436.0,113.0){\rule[-0.200pt]{0.400pt}{173.207pt}}
\put(220.0,832.0){\rule[-0.200pt]{292.934pt}{0.400pt}}
\put(45,472){\makebox(0,0){\shortstack{ $ \frac{ \partial^2 \sigma }{ \partial \tau} $ \\ $ \hbox{ (fbarn)} $ }}}
\put(828,23){\makebox(0,0){ $ \tau $ }}
\put(828,877){\makebox(0,0){Fig.1.}}
\put(220.0,113.0){\rule[-0.200pt]{0.400pt}{173.207pt}}
\put(1306,767){\makebox(0,0)[r]{ $ B^+ B^- , m_B=400 \hbox{GeV },  \sqrt{s}=14 $ TeV}}
\put(1328.0,767.0){\rule[-0.200pt]{15.899pt}{0.400pt}}
\put(240,113){\usebox{\plotpoint}}
\put(240.17,113){\rule{0.400pt}{24.300pt}}
\multiput(239.17,113.00)(2.000,70.564){2}{\rule{0.400pt}{12.150pt}}
\put(241.67,234){\rule{0.400pt}{17.345pt}}
\multiput(241.17,234.00)(1.000,36.000){2}{\rule{0.400pt}{8.672pt}}
\multiput(243.59,306.00)(0.477,29.542){7}{\rule{0.115pt}{21.380pt}}
\multiput(242.17,306.00)(5.000,221.625){2}{\rule{0.400pt}{10.690pt}}
\put(247.67,572){\rule{0.400pt}{7.950pt}}
\multiput(247.17,572.00)(1.000,16.500){2}{\rule{0.400pt}{3.975pt}}
\multiput(249.59,605.00)(0.477,10.283){7}{\rule{0.115pt}{7.540pt}}
\multiput(248.17,605.00)(5.000,77.350){2}{\rule{0.400pt}{3.770pt}}
\put(253.67,698){\rule{0.400pt}{2.168pt}}
\multiput(253.17,698.00)(1.000,4.500){2}{\rule{0.400pt}{1.084pt}}
\multiput(255.59,707.00)(0.477,1.378){7}{\rule{0.115pt}{1.140pt}}
\multiput(254.17,707.00)(5.000,10.634){2}{\rule{0.400pt}{0.570pt}}
\put(260,718.67){\rule{0.241pt}{0.400pt}}
\multiput(260.00,719.17)(0.500,-1.000){2}{\rule{0.120pt}{0.400pt}}
\multiput(261.59,711.94)(0.477,-2.157){7}{\rule{0.115pt}{1.700pt}}
\multiput(260.17,715.47)(5.000,-16.472){2}{\rule{0.400pt}{0.850pt}}
\put(265.67,694){\rule{0.400pt}{1.204pt}}
\multiput(265.17,696.50)(1.000,-2.500){2}{\rule{0.400pt}{0.602pt}}
\multiput(267.59,682.96)(0.477,-3.493){7}{\rule{0.115pt}{2.660pt}}
\multiput(266.17,688.48)(5.000,-26.479){2}{\rule{0.400pt}{1.330pt}}
\put(271.67,656){\rule{0.400pt}{1.445pt}}
\multiput(271.17,659.00)(1.000,-3.000){2}{\rule{0.400pt}{0.723pt}}
\multiput(273.59,643.96)(0.477,-3.827){7}{\rule{0.115pt}{2.900pt}}
\multiput(272.17,649.98)(5.000,-28.981){2}{\rule{0.400pt}{1.450pt}}
\put(277.67,614){\rule{0.400pt}{1.686pt}}
\multiput(277.17,617.50)(1.000,-3.500){2}{\rule{0.400pt}{0.843pt}}
\multiput(279.59,602.52)(0.482,-3.564){9}{\rule{0.116pt}{2.767pt}}
\multiput(278.17,608.26)(6.000,-34.258){2}{\rule{0.400pt}{1.383pt}}
\multiput(285.59,564.57)(0.485,-2.857){11}{\rule{0.117pt}{2.271pt}}
\multiput(284.17,569.29)(7.000,-33.286){2}{\rule{0.400pt}{1.136pt}}
\multiput(292.59,525.90)(0.482,-3.112){9}{\rule{0.116pt}{2.433pt}}
\multiput(291.17,530.95)(6.000,-29.949){2}{\rule{0.400pt}{1.217pt}}
\multiput(298.59,491.73)(0.482,-2.841){9}{\rule{0.116pt}{2.233pt}}
\multiput(297.17,496.36)(6.000,-27.365){2}{\rule{0.400pt}{1.117pt}}
\multiput(304.58,461.11)(0.492,-2.306){21}{\rule{0.119pt}{1.900pt}}
\multiput(303.17,465.06)(12.000,-50.056){2}{\rule{0.400pt}{0.950pt}}
\multiput(316.59,408.22)(0.482,-2.027){9}{\rule{0.116pt}{1.633pt}}
\multiput(315.17,411.61)(6.000,-19.610){2}{\rule{0.400pt}{0.817pt}}
\multiput(322.59,385.77)(0.482,-1.847){9}{\rule{0.116pt}{1.500pt}}
\multiput(321.17,388.89)(6.000,-17.887){2}{\rule{0.400pt}{0.750pt}}
\multiput(328.59,365.33)(0.482,-1.666){9}{\rule{0.116pt}{1.367pt}}
\multiput(327.17,368.16)(6.000,-16.163){2}{\rule{0.400pt}{0.683pt}}
\multiput(334.59,346.88)(0.482,-1.485){9}{\rule{0.116pt}{1.233pt}}
\multiput(333.17,349.44)(6.000,-14.440){2}{\rule{0.400pt}{0.617pt}}
\multiput(340.59,330.16)(0.482,-1.395){9}{\rule{0.116pt}{1.167pt}}
\multiput(339.17,332.58)(6.000,-13.579){2}{\rule{0.400pt}{0.583pt}}
\multiput(346.58,314.85)(0.492,-1.142){21}{\rule{0.119pt}{1.000pt}}
\multiput(345.17,316.92)(12.000,-24.924){2}{\rule{0.400pt}{0.500pt}}
\multiput(358.59,288.26)(0.482,-1.033){9}{\rule{0.116pt}{0.900pt}}
\multiput(357.17,290.13)(6.000,-10.132){2}{\rule{0.400pt}{0.450pt}}
\multiput(364.59,277.21)(0.485,-0.721){11}{\rule{0.117pt}{0.671pt}}
\multiput(363.17,278.61)(7.000,-8.606){2}{\rule{0.400pt}{0.336pt}}
\multiput(371.59,266.82)(0.482,-0.852){9}{\rule{0.116pt}{0.767pt}}
\multiput(370.17,268.41)(6.000,-8.409){2}{\rule{0.400pt}{0.383pt}}
\multiput(377.59,257.09)(0.482,-0.762){9}{\rule{0.116pt}{0.700pt}}
\multiput(376.17,258.55)(6.000,-7.547){2}{\rule{0.400pt}{0.350pt}}
\multiput(383.59,248.09)(0.482,-0.762){9}{\rule{0.116pt}{0.700pt}}
\multiput(382.17,249.55)(6.000,-7.547){2}{\rule{0.400pt}{0.350pt}}
\multiput(389.58,239.51)(0.492,-0.625){21}{\rule{0.119pt}{0.600pt}}
\multiput(388.17,240.75)(12.000,-13.755){2}{\rule{0.400pt}{0.300pt}}
\multiput(401.59,224.65)(0.482,-0.581){9}{\rule{0.116pt}{0.567pt}}
\multiput(400.17,225.82)(6.000,-5.824){2}{\rule{0.400pt}{0.283pt}}
\multiput(407.00,218.93)(0.491,-0.482){9}{\rule{0.500pt}{0.116pt}}
\multiput(407.00,219.17)(4.962,-6.000){2}{\rule{0.250pt}{0.400pt}}
\multiput(413.00,212.93)(0.599,-0.477){7}{\rule{0.580pt}{0.115pt}}
\multiput(413.00,213.17)(4.796,-5.000){2}{\rule{0.290pt}{0.400pt}}
\multiput(419.00,207.93)(0.491,-0.482){9}{\rule{0.500pt}{0.116pt}}
\multiput(419.00,208.17)(4.962,-6.000){2}{\rule{0.250pt}{0.400pt}}
\multiput(425.00,201.93)(0.599,-0.477){7}{\rule{0.580pt}{0.115pt}}
\multiput(425.00,202.17)(4.796,-5.000){2}{\rule{0.290pt}{0.400pt}}
\multiput(431.00,196.93)(0.728,-0.489){15}{\rule{0.678pt}{0.118pt}}
\multiput(431.00,197.17)(11.593,-9.000){2}{\rule{0.339pt}{0.400pt}}
\multiput(444.00,187.94)(0.774,-0.468){5}{\rule{0.700pt}{0.113pt}}
\multiput(444.00,188.17)(4.547,-4.000){2}{\rule{0.350pt}{0.400pt}}
\multiput(450.00,183.95)(1.132,-0.447){3}{\rule{0.900pt}{0.108pt}}
\multiput(450.00,184.17)(4.132,-3.000){2}{\rule{0.450pt}{0.400pt}}
\multiput(456.00,180.94)(0.774,-0.468){5}{\rule{0.700pt}{0.113pt}}
\multiput(456.00,181.17)(4.547,-4.000){2}{\rule{0.350pt}{0.400pt}}
\multiput(462.00,176.95)(1.132,-0.447){3}{\rule{0.900pt}{0.108pt}}
\multiput(462.00,177.17)(4.132,-3.000){2}{\rule{0.450pt}{0.400pt}}
\multiput(468.00,173.95)(1.132,-0.447){3}{\rule{0.900pt}{0.108pt}}
\multiput(468.00,174.17)(4.132,-3.000){2}{\rule{0.450pt}{0.400pt}}
\multiput(474.00,170.93)(1.033,-0.482){9}{\rule{0.900pt}{0.116pt}}
\multiput(474.00,171.17)(10.132,-6.000){2}{\rule{0.450pt}{0.400pt}}
\multiput(486.00,164.95)(1.132,-0.447){3}{\rule{0.900pt}{0.108pt}}
\multiput(486.00,165.17)(4.132,-3.000){2}{\rule{0.450pt}{0.400pt}}
\put(492,161.17){\rule{1.300pt}{0.400pt}}
\multiput(492.00,162.17)(3.302,-2.000){2}{\rule{0.650pt}{0.400pt}}
\put(498,159.17){\rule{1.300pt}{0.400pt}}
\multiput(498.00,160.17)(3.302,-2.000){2}{\rule{0.650pt}{0.400pt}}
\multiput(504.00,157.95)(1.132,-0.447){3}{\rule{0.900pt}{0.108pt}}
\multiput(504.00,158.17)(4.132,-3.000){2}{\rule{0.450pt}{0.400pt}}
\put(510,154.17){\rule{1.300pt}{0.400pt}}
\multiput(510.00,155.17)(3.302,-2.000){2}{\rule{0.650pt}{0.400pt}}
\multiput(516.00,152.95)(2.695,-0.447){3}{\rule{1.833pt}{0.108pt}}
\multiput(516.00,153.17)(9.195,-3.000){2}{\rule{0.917pt}{0.400pt}}
\put(529,149.17){\rule{1.300pt}{0.400pt}}
\multiput(529.00,150.17)(3.302,-2.000){2}{\rule{0.650pt}{0.400pt}}
\put(535,147.17){\rule{1.300pt}{0.400pt}}
\multiput(535.00,148.17)(3.302,-2.000){2}{\rule{0.650pt}{0.400pt}}
\put(541,145.67){\rule{1.445pt}{0.400pt}}
\multiput(541.00,146.17)(3.000,-1.000){2}{\rule{0.723pt}{0.400pt}}
\put(547,144.17){\rule{1.300pt}{0.400pt}}
\multiput(547.00,145.17)(3.302,-2.000){2}{\rule{0.650pt}{0.400pt}}
\put(553,142.67){\rule{1.445pt}{0.400pt}}
\multiput(553.00,143.17)(3.000,-1.000){2}{\rule{0.723pt}{0.400pt}}
\multiput(559.00,141.95)(2.472,-0.447){3}{\rule{1.700pt}{0.108pt}}
\multiput(559.00,142.17)(8.472,-3.000){2}{\rule{0.850pt}{0.400pt}}
\put(571,138.67){\rule{1.445pt}{0.400pt}}
\multiput(571.00,139.17)(3.000,-1.000){2}{\rule{0.723pt}{0.400pt}}
\put(577,137.17){\rule{2.500pt}{0.400pt}}
\multiput(577.00,138.17)(6.811,-2.000){2}{\rule{1.250pt}{0.400pt}}
\put(589,135.67){\rule{1.686pt}{0.400pt}}
\multiput(589.00,136.17)(3.500,-1.000){2}{\rule{0.843pt}{0.400pt}}
\put(596,134.67){\rule{1.445pt}{0.400pt}}
\multiput(596.00,135.17)(3.000,-1.000){2}{\rule{0.723pt}{0.400pt}}
\put(602,133.17){\rule{2.500pt}{0.400pt}}
\multiput(602.00,134.17)(6.811,-2.000){2}{\rule{1.250pt}{0.400pt}}
\put(614,131.67){\rule{1.445pt}{0.400pt}}
\multiput(614.00,132.17)(3.000,-1.000){2}{\rule{0.723pt}{0.400pt}}
\put(620,130.67){\rule{2.891pt}{0.400pt}}
\multiput(620.00,131.17)(6.000,-1.000){2}{\rule{1.445pt}{0.400pt}}
\put(632,129.67){\rule{1.445pt}{0.400pt}}
\multiput(632.00,130.17)(3.000,-1.000){2}{\rule{0.723pt}{0.400pt}}
\put(638,128.17){\rule{3.700pt}{0.400pt}}
\multiput(638.00,129.17)(10.320,-2.000){2}{\rule{1.850pt}{0.400pt}}
\put(656,126.67){\rule{2.891pt}{0.400pt}}
\multiput(656.00,127.17)(6.000,-1.000){2}{\rule{1.445pt}{0.400pt}}
\put(668,125.67){\rule{1.686pt}{0.400pt}}
\multiput(668.00,126.17)(3.500,-1.000){2}{\rule{0.843pt}{0.400pt}}
\put(681,124.67){\rule{1.445pt}{0.400pt}}
\multiput(681.00,125.17)(3.000,-1.000){2}{\rule{0.723pt}{0.400pt}}
\put(687,123.67){\rule{2.891pt}{0.400pt}}
\multiput(687.00,124.17)(6.000,-1.000){2}{\rule{1.445pt}{0.400pt}}
\put(675.0,126.0){\rule[-0.200pt]{1.445pt}{0.400pt}}
\put(705,122.67){\rule{2.891pt}{0.400pt}}
\multiput(705.00,123.17)(6.000,-1.000){2}{\rule{1.445pt}{0.400pt}}
\put(717,121.67){\rule{1.445pt}{0.400pt}}
\multiput(717.00,122.17)(3.000,-1.000){2}{\rule{0.723pt}{0.400pt}}
\put(723,120.67){\rule{4.336pt}{0.400pt}}
\multiput(723.00,121.17)(9.000,-1.000){2}{\rule{2.168pt}{0.400pt}}
\put(741,119.67){\rule{4.577pt}{0.400pt}}
\multiput(741.00,120.17)(9.500,-1.000){2}{\rule{2.289pt}{0.400pt}}
\put(699.0,124.0){\rule[-0.200pt]{1.445pt}{0.400pt}}
\put(772,118.67){\rule{2.891pt}{0.400pt}}
\multiput(772.00,119.17)(6.000,-1.000){2}{\rule{1.445pt}{0.400pt}}
\put(760.0,120.0){\rule[-0.200pt]{2.891pt}{0.400pt}}
\put(808,117.67){\rule{1.445pt}{0.400pt}}
\multiput(808.00,118.17)(3.000,-1.000){2}{\rule{0.723pt}{0.400pt}}
\put(784.0,119.0){\rule[-0.200pt]{5.782pt}{0.400pt}}
\put(839,116.67){\rule{5.782pt}{0.400pt}}
\multiput(839.00,117.17)(12.000,-1.000){2}{\rule{2.891pt}{0.400pt}}
\put(814.0,118.0){\rule[-0.200pt]{6.022pt}{0.400pt}}
\put(875,115.67){\rule{4.336pt}{0.400pt}}
\multiput(875.00,116.17)(9.000,-1.000){2}{\rule{2.168pt}{0.400pt}}
\put(863.0,117.0){\rule[-0.200pt]{2.891pt}{0.400pt}}
\put(930,114.67){\rule{4.336pt}{0.400pt}}
\multiput(930.00,115.17)(9.000,-1.000){2}{\rule{2.168pt}{0.400pt}}
\put(893.0,116.0){\rule[-0.200pt]{8.913pt}{0.400pt}}
\put(1009,113.67){\rule{8.672pt}{0.400pt}}
\multiput(1009.00,114.17)(18.000,-1.000){2}{\rule{4.336pt}{0.400pt}}
\put(948.0,115.0){\rule[-0.200pt]{14.695pt}{0.400pt}}
\put(1191,112.67){\rule{23.608pt}{0.400pt}}
\multiput(1191.00,113.17)(49.000,-1.000){2}{\rule{11.804pt}{0.400pt}}
\put(1045.0,114.0){\rule[-0.200pt]{35.171pt}{0.400pt}}
\put(1289.0,113.0){\rule[-0.200pt]{35.412pt}{0.400pt}}
\end{picture}

\vspace*{2cm}

\setlength{\unitlength}{0.240900pt}
\ifx\plotpoint\undefined\newsavebox{\plotpoint}\fi
\sbox{\plotpoint}{\rule[-0.200pt]{0.400pt}{0.400pt}}%
\begin{picture}(1500,900)(0,0)
\font\gnuplot=cmr10 at 10pt
\gnuplot
\sbox{\plotpoint}{\rule[-0.200pt]{0.400pt}{0.400pt}}%
\put(220.0,113.0){\rule[-0.200pt]{292.934pt}{0.400pt}}
\put(220.0,113.0){\rule[-0.200pt]{0.400pt}{173.207pt}}
\put(220.0,113.0){\rule[-0.200pt]{4.818pt}{0.400pt}}
\put(198,113){\makebox(0,0)[r]{0}}
\put(1416.0,113.0){\rule[-0.200pt]{4.818pt}{0.400pt}}
\put(220.0,257.0){\rule[-0.200pt]{4.818pt}{0.400pt}}
\put(198,257){\makebox(0,0)[r]{0.05}}
\put(1416.0,257.0){\rule[-0.200pt]{4.818pt}{0.400pt}}
\put(220.0,401.0){\rule[-0.200pt]{4.818pt}{0.400pt}}
\put(198,401){\makebox(0,0)[r]{0.1}}
\put(1416.0,401.0){\rule[-0.200pt]{4.818pt}{0.400pt}}
\put(220.0,544.0){\rule[-0.200pt]{4.818pt}{0.400pt}}
\put(198,544){\makebox(0,0)[r]{0.15}}
\put(1416.0,544.0){\rule[-0.200pt]{4.818pt}{0.400pt}}
\put(220.0,688.0){\rule[-0.200pt]{4.818pt}{0.400pt}}
\put(198,688){\makebox(0,0)[r]{0.2}}
\put(1416.0,688.0){\rule[-0.200pt]{4.818pt}{0.400pt}}
\put(220.0,832.0){\rule[-0.200pt]{4.818pt}{0.400pt}}
\put(198,832){\makebox(0,0)[r]{0.25}}
\put(1416.0,832.0){\rule[-0.200pt]{4.818pt}{0.400pt}}
\put(220.0,113.0){\rule[-0.200pt]{0.400pt}{4.818pt}}
\put(220,68){\makebox(0,0){0}}
\put(220.0,812.0){\rule[-0.200pt]{0.400pt}{4.818pt}}
\put(394.0,113.0){\rule[-0.200pt]{0.400pt}{4.818pt}}
\put(394,68){\makebox(0,0){500}}
\put(394.0,812.0){\rule[-0.200pt]{0.400pt}{4.818pt}}
\put(567.0,113.0){\rule[-0.200pt]{0.400pt}{4.818pt}}
\put(567,68){\makebox(0,0){1000}}
\put(567.0,812.0){\rule[-0.200pt]{0.400pt}{4.818pt}}
\put(741.0,113.0){\rule[-0.200pt]{0.400pt}{4.818pt}}
\put(741,68){\makebox(0,0){1500}}
\put(741.0,812.0){\rule[-0.200pt]{0.400pt}{4.818pt}}
\put(915.0,113.0){\rule[-0.200pt]{0.400pt}{4.818pt}}
\put(915,68){\makebox(0,0){2000}}
\put(915.0,812.0){\rule[-0.200pt]{0.400pt}{4.818pt}}
\put(1089.0,113.0){\rule[-0.200pt]{0.400pt}{4.818pt}}
\put(1089,68){\makebox(0,0){2500}}
\put(1089.0,812.0){\rule[-0.200pt]{0.400pt}{4.818pt}}
\put(1262.0,113.0){\rule[-0.200pt]{0.400pt}{4.818pt}}
\put(1262,68){\makebox(0,0){3000}}
\put(1262.0,812.0){\rule[-0.200pt]{0.400pt}{4.818pt}}
\put(1436.0,113.0){\rule[-0.200pt]{0.400pt}{4.818pt}}
\put(1436,68){\makebox(0,0){3500}}
\put(1436.0,812.0){\rule[-0.200pt]{0.400pt}{4.818pt}}
\put(220.0,113.0){\rule[-0.200pt]{292.934pt}{0.400pt}}
\put(1436.0,113.0){\rule[-0.200pt]{0.400pt}{173.207pt}}
\put(220.0,832.0){\rule[-0.200pt]{292.934pt}{0.400pt}}
\put(-21,472){\makebox(0,0){\shortstack{ $ \frac{ \partial \sigma }{ \partial p_T \partial y} \vert_{y=0} $ \\ $ (\hbox{ fbarn GeV}^{-1}) $ }}}
\put(828,23){\makebox(0,0){ $ p_T $  (GeV) }}
\put(828,877){\makebox(0,0){Fig.2.}}
\put(220.0,113.0){\rule[-0.200pt]{0.400pt}{173.207pt}}
\put(1306,767){\makebox(0,0)[r]{ $ B^+ B^-,  m_B=400 \hbox{GeV},  \sqrt{s}=14 $ TeV}}
\put(1328.0,767.0){\rule[-0.200pt]{15.899pt}{0.400pt}}
\put(220,113){\usebox{\plotpoint}}
\multiput(220.59,113.00)(0.482,3.293){9}{\rule{0.116pt}{2.567pt}}
\multiput(219.17,113.00)(6.000,31.673){2}{\rule{0.400pt}{1.283pt}}
\multiput(226.58,150.00)(0.492,3.210){21}{\rule{0.119pt}{2.600pt}}
\multiput(225.17,150.00)(12.000,69.604){2}{\rule{0.400pt}{1.300pt}}
\multiput(238.58,225.00)(0.492,3.124){21}{\rule{0.119pt}{2.533pt}}
\multiput(237.17,225.00)(12.000,67.742){2}{\rule{0.400pt}{1.267pt}}
\multiput(250.58,298.00)(0.492,2.995){21}{\rule{0.119pt}{2.433pt}}
\multiput(249.17,298.00)(12.000,64.949){2}{\rule{0.400pt}{1.217pt}}
\multiput(262.58,368.00)(0.493,2.519){23}{\rule{0.119pt}{2.069pt}}
\multiput(261.17,368.00)(13.000,59.705){2}{\rule{0.400pt}{1.035pt}}
\multiput(275.58,432.00)(0.492,2.693){21}{\rule{0.119pt}{2.200pt}}
\multiput(274.17,432.00)(12.000,58.434){2}{\rule{0.400pt}{1.100pt}}
\multiput(287.58,495.00)(0.492,2.392){21}{\rule{0.119pt}{1.967pt}}
\multiput(286.17,495.00)(12.000,51.918){2}{\rule{0.400pt}{0.983pt}}
\multiput(299.58,551.00)(0.492,2.090){21}{\rule{0.119pt}{1.733pt}}
\multiput(298.17,551.00)(12.000,45.402){2}{\rule{0.400pt}{0.867pt}}
\multiput(311.58,600.00)(0.492,1.789){21}{\rule{0.119pt}{1.500pt}}
\multiput(310.17,600.00)(12.000,38.887){2}{\rule{0.400pt}{0.750pt}}
\multiput(323.58,642.00)(0.492,1.487){21}{\rule{0.119pt}{1.267pt}}
\multiput(322.17,642.00)(12.000,32.371){2}{\rule{0.400pt}{0.633pt}}
\multiput(335.58,677.00)(0.492,1.142){21}{\rule{0.119pt}{1.000pt}}
\multiput(334.17,677.00)(12.000,24.924){2}{\rule{0.400pt}{0.500pt}}
\multiput(347.58,704.00)(0.493,0.774){23}{\rule{0.119pt}{0.715pt}}
\multiput(346.17,704.00)(13.000,18.515){2}{\rule{0.400pt}{0.358pt}}
\multiput(360.00,724.58)(0.496,0.492){21}{\rule{0.500pt}{0.119pt}}
\multiput(360.00,723.17)(10.962,12.000){2}{\rule{0.250pt}{0.400pt}}
\multiput(372.00,736.59)(1.033,0.482){9}{\rule{0.900pt}{0.116pt}}
\multiput(372.00,735.17)(10.132,6.000){2}{\rule{0.450pt}{0.400pt}}
\multiput(396.00,740.93)(1.033,-0.482){9}{\rule{0.900pt}{0.116pt}}
\multiput(396.00,741.17)(10.132,-6.000){2}{\rule{0.450pt}{0.400pt}}
\multiput(408.00,734.92)(0.543,-0.492){19}{\rule{0.536pt}{0.118pt}}
\multiput(408.00,735.17)(10.887,-11.000){2}{\rule{0.268pt}{0.400pt}}
\multiput(420.58,722.65)(0.492,-0.582){21}{\rule{0.119pt}{0.567pt}}
\multiput(419.17,723.82)(12.000,-12.824){2}{\rule{0.400pt}{0.283pt}}
\multiput(432.58,708.29)(0.493,-0.695){23}{\rule{0.119pt}{0.654pt}}
\multiput(431.17,709.64)(13.000,-16.643){2}{\rule{0.400pt}{0.327pt}}
\multiput(445.58,689.82)(0.492,-0.841){21}{\rule{0.119pt}{0.767pt}}
\multiput(444.17,691.41)(12.000,-18.409){2}{\rule{0.400pt}{0.383pt}}
\multiput(457.58,669.40)(0.492,-0.970){21}{\rule{0.119pt}{0.867pt}}
\multiput(456.17,671.20)(12.000,-21.201){2}{\rule{0.400pt}{0.433pt}}
\multiput(469.58,646.40)(0.492,-0.970){21}{\rule{0.119pt}{0.867pt}}
\multiput(468.17,648.20)(12.000,-21.201){2}{\rule{0.400pt}{0.433pt}}
\multiput(481.58,623.13)(0.492,-1.056){21}{\rule{0.119pt}{0.933pt}}
\multiput(480.17,625.06)(12.000,-23.063){2}{\rule{0.400pt}{0.467pt}}
\multiput(493.58,598.13)(0.492,-1.056){21}{\rule{0.119pt}{0.933pt}}
\multiput(492.17,600.06)(12.000,-23.063){2}{\rule{0.400pt}{0.467pt}}
\multiput(505.58,572.99)(0.492,-1.099){21}{\rule{0.119pt}{0.967pt}}
\multiput(504.17,574.99)(12.000,-23.994){2}{\rule{0.400pt}{0.483pt}}
\multiput(517.58,547.39)(0.493,-0.972){23}{\rule{0.119pt}{0.869pt}}
\multiput(516.17,549.20)(13.000,-23.196){2}{\rule{0.400pt}{0.435pt}}
\multiput(530.58,522.13)(0.492,-1.056){21}{\rule{0.119pt}{0.933pt}}
\multiput(529.17,524.06)(12.000,-23.063){2}{\rule{0.400pt}{0.467pt}}
\multiput(542.58,497.13)(0.492,-1.056){21}{\rule{0.119pt}{0.933pt}}
\multiput(541.17,499.06)(12.000,-23.063){2}{\rule{0.400pt}{0.467pt}}
\multiput(554.58,472.40)(0.492,-0.970){21}{\rule{0.119pt}{0.867pt}}
\multiput(553.17,474.20)(12.000,-21.201){2}{\rule{0.400pt}{0.433pt}}
\multiput(566.58,449.40)(0.492,-0.970){21}{\rule{0.119pt}{0.867pt}}
\multiput(565.17,451.20)(12.000,-21.201){2}{\rule{0.400pt}{0.433pt}}
\multiput(578.58,426.54)(0.492,-0.927){21}{\rule{0.119pt}{0.833pt}}
\multiput(577.17,428.27)(12.000,-20.270){2}{\rule{0.400pt}{0.417pt}}
\multiput(590.58,404.68)(0.492,-0.884){21}{\rule{0.119pt}{0.800pt}}
\multiput(589.17,406.34)(12.000,-19.340){2}{\rule{0.400pt}{0.400pt}}
\multiput(602.58,384.03)(0.493,-0.774){23}{\rule{0.119pt}{0.715pt}}
\multiput(601.17,385.52)(13.000,-18.515){2}{\rule{0.400pt}{0.358pt}}
\multiput(615.58,363.96)(0.492,-0.798){21}{\rule{0.119pt}{0.733pt}}
\multiput(614.17,365.48)(12.000,-17.478){2}{\rule{0.400pt}{0.367pt}}
\multiput(627.58,345.09)(0.492,-0.755){21}{\rule{0.119pt}{0.700pt}}
\multiput(626.17,346.55)(12.000,-16.547){2}{\rule{0.400pt}{0.350pt}}
\multiput(639.58,327.23)(0.492,-0.712){21}{\rule{0.119pt}{0.667pt}}
\multiput(638.17,328.62)(12.000,-15.616){2}{\rule{0.400pt}{0.333pt}}
\multiput(651.58,310.37)(0.492,-0.669){21}{\rule{0.119pt}{0.633pt}}
\multiput(650.17,311.69)(12.000,-14.685){2}{\rule{0.400pt}{0.317pt}}
\multiput(663.58,294.65)(0.492,-0.582){21}{\rule{0.119pt}{0.567pt}}
\multiput(662.17,295.82)(12.000,-12.824){2}{\rule{0.400pt}{0.283pt}}
\multiput(675.58,280.65)(0.492,-0.582){21}{\rule{0.119pt}{0.567pt}}
\multiput(674.17,281.82)(12.000,-12.824){2}{\rule{0.400pt}{0.283pt}}
\multiput(687.00,267.92)(0.497,-0.493){23}{\rule{0.500pt}{0.119pt}}
\multiput(687.00,268.17)(11.962,-13.000){2}{\rule{0.250pt}{0.400pt}}
\multiput(700.00,254.92)(0.496,-0.492){21}{\rule{0.500pt}{0.119pt}}
\multiput(700.00,255.17)(10.962,-12.000){2}{\rule{0.250pt}{0.400pt}}
\multiput(712.00,242.92)(0.543,-0.492){19}{\rule{0.536pt}{0.118pt}}
\multiput(712.00,243.17)(10.887,-11.000){2}{\rule{0.268pt}{0.400pt}}
\multiput(724.00,231.92)(0.600,-0.491){17}{\rule{0.580pt}{0.118pt}}
\multiput(724.00,232.17)(10.796,-10.000){2}{\rule{0.290pt}{0.400pt}}
\multiput(736.00,221.92)(0.600,-0.491){17}{\rule{0.580pt}{0.118pt}}
\multiput(736.00,222.17)(10.796,-10.000){2}{\rule{0.290pt}{0.400pt}}
\multiput(748.00,211.93)(0.669,-0.489){15}{\rule{0.633pt}{0.118pt}}
\multiput(748.00,212.17)(10.685,-9.000){2}{\rule{0.317pt}{0.400pt}}
\multiput(760.00,202.93)(0.758,-0.488){13}{\rule{0.700pt}{0.117pt}}
\multiput(760.00,203.17)(10.547,-8.000){2}{\rule{0.350pt}{0.400pt}}
\multiput(772.00,194.93)(0.950,-0.485){11}{\rule{0.843pt}{0.117pt}}
\multiput(772.00,195.17)(11.251,-7.000){2}{\rule{0.421pt}{0.400pt}}
\multiput(785.00,187.93)(0.874,-0.485){11}{\rule{0.786pt}{0.117pt}}
\multiput(785.00,188.17)(10.369,-7.000){2}{\rule{0.393pt}{0.400pt}}
\multiput(797.00,180.93)(0.874,-0.485){11}{\rule{0.786pt}{0.117pt}}
\multiput(797.00,181.17)(10.369,-7.000){2}{\rule{0.393pt}{0.400pt}}
\multiput(809.00,173.93)(1.267,-0.477){7}{\rule{1.060pt}{0.115pt}}
\multiput(809.00,174.17)(9.800,-5.000){2}{\rule{0.530pt}{0.400pt}}
\multiput(821.00,168.93)(1.033,-0.482){9}{\rule{0.900pt}{0.116pt}}
\multiput(821.00,169.17)(10.132,-6.000){2}{\rule{0.450pt}{0.400pt}}
\multiput(833.00,162.93)(1.267,-0.477){7}{\rule{1.060pt}{0.115pt}}
\multiput(833.00,163.17)(9.800,-5.000){2}{\rule{0.530pt}{0.400pt}}
\multiput(845.00,157.94)(1.651,-0.468){5}{\rule{1.300pt}{0.113pt}}
\multiput(845.00,158.17)(9.302,-4.000){2}{\rule{0.650pt}{0.400pt}}
\multiput(857.00,153.94)(1.651,-0.468){5}{\rule{1.300pt}{0.113pt}}
\multiput(857.00,154.17)(9.302,-4.000){2}{\rule{0.650pt}{0.400pt}}
\multiput(869.00,149.94)(1.797,-0.468){5}{\rule{1.400pt}{0.113pt}}
\multiput(869.00,150.17)(10.094,-4.000){2}{\rule{0.700pt}{0.400pt}}
\multiput(882.00,145.95)(2.472,-0.447){3}{\rule{1.700pt}{0.108pt}}
\multiput(882.00,146.17)(8.472,-3.000){2}{\rule{0.850pt}{0.400pt}}
\multiput(894.00,142.95)(2.472,-0.447){3}{\rule{1.700pt}{0.108pt}}
\multiput(894.00,143.17)(8.472,-3.000){2}{\rule{0.850pt}{0.400pt}}
\multiput(906.00,139.95)(2.472,-0.447){3}{\rule{1.700pt}{0.108pt}}
\multiput(906.00,140.17)(8.472,-3.000){2}{\rule{0.850pt}{0.400pt}}
\multiput(918.00,136.95)(2.472,-0.447){3}{\rule{1.700pt}{0.108pt}}
\multiput(918.00,137.17)(8.472,-3.000){2}{\rule{0.850pt}{0.400pt}}
\multiput(930.00,133.94)(3.406,-0.468){5}{\rule{2.500pt}{0.113pt}}
\multiput(930.00,134.17)(18.811,-4.000){2}{\rule{1.250pt}{0.400pt}}
\put(954,129.17){\rule{2.700pt}{0.400pt}}
\multiput(954.00,130.17)(7.396,-2.000){2}{\rule{1.350pt}{0.400pt}}
\put(967,127.17){\rule{2.500pt}{0.400pt}}
\multiput(967.00,128.17)(6.811,-2.000){2}{\rule{1.250pt}{0.400pt}}
\put(979,125.67){\rule{2.891pt}{0.400pt}}
\multiput(979.00,126.17)(6.000,-1.000){2}{\rule{1.445pt}{0.400pt}}
\put(991,124.67){\rule{2.891pt}{0.400pt}}
\multiput(991.00,125.17)(6.000,-1.000){2}{\rule{1.445pt}{0.400pt}}
\put(1003,123.17){\rule{2.500pt}{0.400pt}}
\multiput(1003.00,124.17)(6.811,-2.000){2}{\rule{1.250pt}{0.400pt}}
\put(1015,121.67){\rule{2.891pt}{0.400pt}}
\multiput(1015.00,122.17)(6.000,-1.000){2}{\rule{1.445pt}{0.400pt}}
\put(1027,120.67){\rule{2.891pt}{0.400pt}}
\multiput(1027.00,121.17)(6.000,-1.000){2}{\rule{1.445pt}{0.400pt}}
\put(1039,119.67){\rule{3.132pt}{0.400pt}}
\multiput(1039.00,120.17)(6.500,-1.000){2}{\rule{1.566pt}{0.400pt}}
\put(1052,118.67){\rule{2.891pt}{0.400pt}}
\multiput(1052.00,119.17)(6.000,-1.000){2}{\rule{1.445pt}{0.400pt}}
\put(384.0,742.0){\rule[-0.200pt]{2.891pt}{0.400pt}}
\put(1076,117.67){\rule{2.891pt}{0.400pt}}
\multiput(1076.00,118.17)(6.000,-1.000){2}{\rule{1.445pt}{0.400pt}}
\put(1088,116.67){\rule{2.891pt}{0.400pt}}
\multiput(1088.00,117.17)(6.000,-1.000){2}{\rule{1.445pt}{0.400pt}}
\put(1100,115.67){\rule{5.782pt}{0.400pt}}
\multiput(1100.00,116.17)(12.000,-1.000){2}{\rule{2.891pt}{0.400pt}}
\put(1064.0,119.0){\rule[-0.200pt]{2.891pt}{0.400pt}}
\put(1149,114.67){\rule{2.891pt}{0.400pt}}
\multiput(1149.00,115.17)(6.000,-1.000){2}{\rule{1.445pt}{0.400pt}}
\put(1124.0,116.0){\rule[-0.200pt]{6.022pt}{0.400pt}}
\put(1185,113.67){\rule{5.782pt}{0.400pt}}
\multiput(1185.00,114.17)(12.000,-1.000){2}{\rule{2.891pt}{0.400pt}}
\put(1161.0,115.0){\rule[-0.200pt]{5.782pt}{0.400pt}}
\put(1294,112.67){\rule{6.023pt}{0.400pt}}
\multiput(1294.00,113.17)(12.500,-1.000){2}{\rule{3.011pt}{0.400pt}}
\put(1209.0,114.0){\rule[-0.200pt]{20.476pt}{0.400pt}}
\put(1319.0,113.0){\rule[-0.200pt]{26.258pt}{0.400pt}}
\end{picture}

 
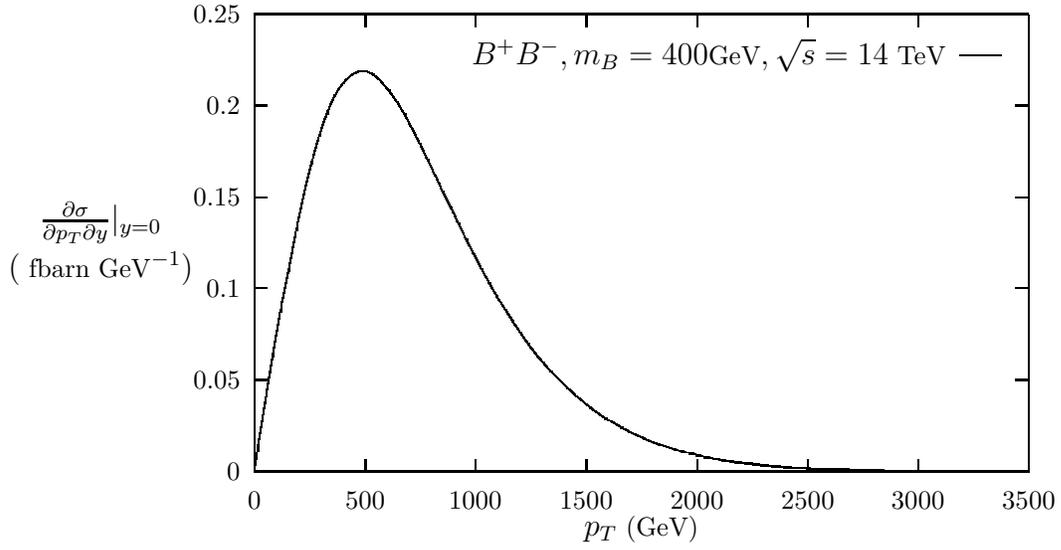
\end{document}